\documentclass{elsart}

\usepackage{amssymb}
\usepackage{epsfig}

\begin{document}

\begin{frontmatter}


% \author{Name\corauthref{CDR1}\thanksref{label2}}

\title{A luminosity monitor for the A4 parity violation experiment at MAMI}

\author{T.~Hammel\corauthref{doc}},
\author{P.~Achenbach},
\author{S.~Baunack},
\author{L.~Capozza},
\author{J.~Diefenbach},
\author{K.~Grimm},
\author{D.~von Harrach},
\author{Y.~Imai},
\author{E.~Kabu{\ss}},
\author{R.~Kothe},
\author{J.~H.~Lee},
\author{A.~Lopes Ginja},
\author{F.~E.~Maas\corauthref{cor}},
\author{A.~Sanchez~Lorente},
\author{E.~Schilling},
\author{G.~Stephan},
\author{C.~Weinrich}
\corauth[doc]{Part of doctoral thesis}
\corauth[cor]{Corresponding author. E-Mail: maas@kph.uni-mainz.de}
\address{Institut f\"ur Kernphysik, Johannes Gutenberg-Universit\"at, D-55099 Mainz, Germany}

\author{I. Altarev\corauthref{asso}}
\address{Technische Universit{\"a}t M{\"u}nchen, D-85748 M{\"u}nchen, Germany}
\corauth[asso]{Associated Member of the St.~Petersburg Nuclear Physics Institute, Russia}

\begin{abstract}
A water Cherenkov luminosity monitor system with associated electronics has been
developed for the A4 parity violation experiment at MAMI. The detector system
measures the luminosity of the hydrogen target hit by the MAMI electron beam
and monitors the stability of the liquid hydrogen target. Both is required for the
precise study of the count rate asymmetries in the scattering of
longitudinally polarized electrons on unpolarized protons. Any helicity
correlated fluctuation of the target density leads to false
asymmetries. The performance of the luminosity monitor, investigated
in about 2000 hours with electron beam, and the results of its application in the
A4 experiment are presented.
\end{abstract}

\end{frontmatter}

\section{Introduction}
\begin{figure}[htbp]
  \begin{center}
    \epsfig{file=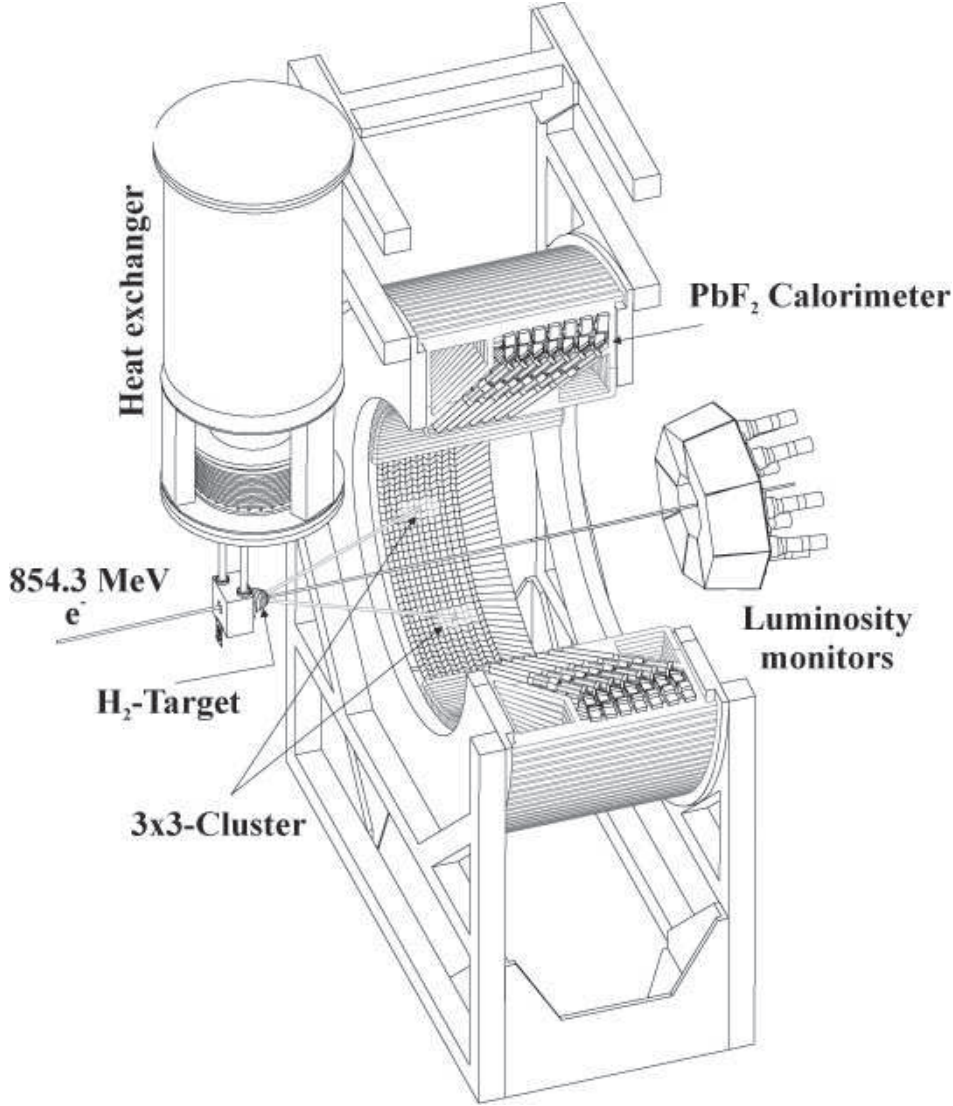, width= 0.9 \textwidth}
    \caption{Set-up of the A4 experiment. The 854.3\,MeV electron beam
      enters the 10~cm liquid hydrogen target cell made from thin aluminum
      from the left. Above the liquid hydrogen
      target the heat exchanger is located. The PbF$_2$ calorimeter
      consists of 1022 crystals and is shown as a sectional
      view with parts of it removed for better visibility.
      The 8 water Cherenkov luminosity monitors are placed at electron scattering angles
      of 4.4$^\circ$ $<$ $\theta_e$ $<$ 10$^\circ$ downstream the calorimeter.}
    \label{fig:a4setup}
  \end{center}
\end{figure}

The A4 collaboration at the MAMI accelerator facility in Mainz
is investigating the contribution of strangeness to the vector form
factors of the nucleon by measuring the weak form factors of the nucleon.
They are accessed
by measuring a parity violating (PV) asymmetry of order $10^{-6}$ in the
cross section of elastic scattering of longitudinally polarized 854.3\,MeV
electrons off unpolarized protons. The scattered electrons are detected by a lead fluoride
(PbF$_2$) calorimeter at electron scattering angles $\theta_e$ of
30$^\circ$ $<$ $\theta_e$ $<$ 40$^\circ$~\cite{maas:pv:2004,maas:2photon:2005,maas:pv:2005,maas:a4cal:2003,maas:a4cal:2002}.
Fig.~\ref{fig:a4setup} shows the
set-up of the A4 experiment. The 854.3\,MeV electron beam enters
the 10~cm liquid hydrogen target cell from the left. The
heat exchanger is located on top of the liquid hydrogen target. The PbF$_2$ calorimeter
is symmetric around the beam axis and consists of 1022
crystals. It is shown as a sectional view.
The eight water Cherenkov luminosity monitors presented in this paper
are placed downstream the calorimeter
at electron scattering angles of 4.4$^\circ$ $<$ $\theta_e$ $<$ 10$^\circ$.
The experimental asymmetry $A_{\mathrm{Exp}}$ is extracted from the number of
counts of elastic scattered electrons $N^{\pm}$ for right ($+$)
and left handed ($-$) electron beam helicity, respectively,
normalized to the integrated effective target density $\rho^{\pm}$:
\begin{eqnarray}
 A_{\mathrm{Exp}}&=&\frac{ (N^+/\rho^+) - (N^-/\rho^-)}{(N^+/\rho^+) + (N^-/\rho^-)} \label{eq:aexp:rho}\\
        &=&\frac{N^+ - N^-}{ N^+ + N^-} + \frac{I^+ - I^-}{I^+ + I^-} - \frac{L^+ - L^-}{L^+ + L^-} + \mathcal{O}(10^{-12})\\
        &=&A_{\mathrm{phys}} + A_I - A_L, \label{eq:aexp:asym}
\end{eqnarray}
where the effective target density $\rho^{\pm}$ of a fixed-target
experiment is the luminosity $L^{\pm}$ divided by the beam current
$I^{\pm}$. $A_{\mathrm{phys}}$ denotes the PV physics asymmetry in
elastic electron proton scattering which is of order
10$^{-6}$.  $A_I$ is a possible apparative asymmetry
in the incoming beam current which is in the experiment of order $10^{-6}$.
$A_L$ corresponds to a possible asymmetry in the luminosity signal also
of order $10^{-6}$.
%The integration time is 20~ms, the data are averaged over
%a 5 minutes run.

The A4 experiment measures very small (order 10$^{-6}$) PV cross section
asymmetries. Accordingly, the following considerations have been taken into
account for the design of the luminosity monitors:
\begin{itemize}
\item[i.] A measurement of the absolute luminosity is not necessary
    since any efficiency or calibration factor cancels in the ratio
    in Eq.~\ref{eq:aexp:rho}, as long as it is exactly equal for both
    helicity states. This is accomplished by flipping the electron beam helicity
    at a rate (25 Hz) which is fast compared to most external fluctuation
    sources \cite{Hammel:dis:2004}.
\item[ii.] The physical process which is used to measure the luminosity
    should have no (or very small) PV asymmetry. Possible processes
    contributing to the measured luminosity signal are
    elastic electron proton scattering at
    small scattering angles and elastic electron electron (M{\o}ller) scattering.
    We have optimized the response behavior of the luminosity monitor
    for the detection of M{\o}ller scattering.
\item[iii.] The accuracy in the PV asymmetry measurement should be
    dominated by the count rate accuracy (i.e. $\sqrt{N}$)
    and not by target density fluctuations,
    which may arise from temperature fluctuations or from
    boiling in the liquid hydrogen target caused by the heat deposition
    of the intense 20~$\mu$A electron beam in the hydrogen target.
    In order to accurately correct the measured asymmetry for target density fluctuations,
    we designed a monitor capable of measuring
    fluctuations with a relative accuracy on the level of $10^{-5}$ in
    20\,ms.
\item[iv.] The correct determination of false asymmetries arising
    from helicity correlated beam parameter fluctuations is necessary
    to measure the luminosity in order to disentangle beam
    current fluctuations from target density fluctuations
    as can be seen in Eq.~\ref{eq:aexp:asym}.
\item[v.] Due to the fact that in the A4 experiment
    counting of individually scattered particles is applied
    to a measurement of a parity violating asymmetry,
    effects of dead time in the counting electronics have to be considered.
    Systematic changes of the measured asymmetry can arise not only
    from differences in the integrated luminosity $L^{\pm}$ for positive ($+$) and
    negative ($-$) helicity, but also from fluctuations or differences of
    fluctuations in the luminosity which makes the simultaneous measurement of the square
    of the luminosity necessary.
\end{itemize}

\section{Luminosity measurement}
In a fixed target experiment the luminosity is given by the product of
the beam current times the area density of scattering centers in the target.
The luminosity of an experiment describes the flux density of the
scattering reaction partners. The product of luminosity $L$ and
cross section $\sigma$ results in an observed reaction rate
$R=\sigma \cdot L$. A possible experimental method to determine the
luminosity is the measurement of the event rate of a scattering
reaction with a well-known cross section. The expected luminosity in the
A4 experiment can be calculated from the flux of the beam particles
$\phi_{e}$ [electrons/s] and the effective area target density $\hat{\rho}_{H_2}$
[atoms/cm$^2$] by $L = \phi_{e} \cdot \hat{\rho}_{H_2}$. For a beam current
of 20\,$\mu$A and a target length of 10\,cm the nominal luminosity in
the A4 experiment is $L=5.3 \cdot 10^{37}$s$^{-1}$~cm$^{-2}$. \\
Target density fluctuations, which are caused by boiling inside the hydrogen volume
of the target and at the entrance and exit aluminum windows, can lead
to large luminosity fluctuations compromising the statistical
accuracy of the experiment.
Since the A4 experiment does not use rastering of the 20\,$\mu$A electron beam
the avoidance of target boiling poses a delicate problem which is
overcome by a highly turbulent flow of liquid hydrogen in the target cell
and a wide electron beam profile in the hydrogen target cell~\cite{Altarev2001}.\\
\begin{figure}[htbp]
  \begin{center}
    \epsfig{file=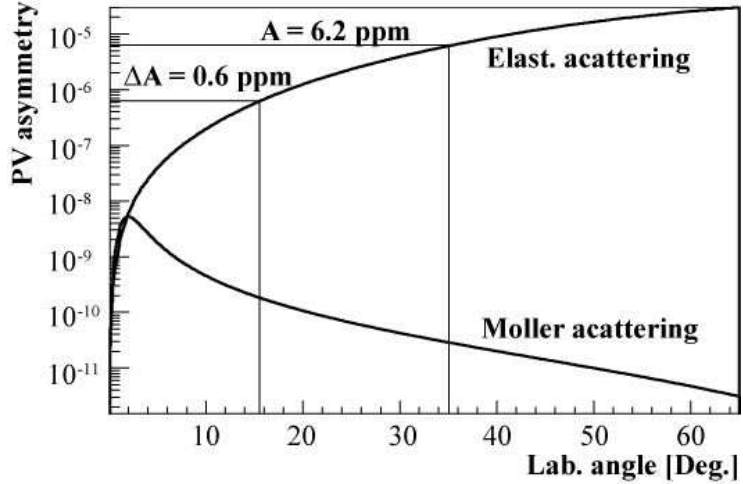,width=0.7\textwidth}
    \caption{Parity violating asymmetries of M{\o}ller scattering and elastic
    scattering as a function of the electron scattering angle in the laboratory
    $\theta_e$.
    The water Cherenkov luminosity detector spans an interval in $\theta_e$
    from $4.4^\circ < \theta_e < 10^\circ$ and the PbF$_2$ calorimeter
    spans an interval from $30^\circ < \theta_e < 40^\circ$.}
    \label{fig:asyma0}
  \end{center}
\end{figure}
Elastic as well as M{\o}ller scattering both exhibit a PV asymmetry. A PV
asymmetry smaller than the statistical accuracy of the A4 experiment is
required for the luminosity measurement in order to be able to
correct for asymmetries of the luminosity with sufficient accuracy.
The asymmetry of elastic and M{\o}ller
scattering is shown in Fig.~\ref{fig:asyma0} as a function of the
electron scattering angle $\theta_e$. For scattering angles $\theta_e \leq 15^\circ$ the asymmetry
in the elastic scattering is smaller than the statistic accuracy of
the A4 experiment ($\delta A_0 = 0.6 \times 10^{-6}$).
In the scattering angle range of the water Cherenkov luminosity detector of
$4.4^\circ < \theta_e < 10^\circ$ the elastic scattering has an asymmetry which is
almost one order of magnitude less than the statistical accuracy of the A4 experiment.
The asymmetry of the dominating M{\o}ller scattering is smaller by two
orders of magnitude. At these
angles the mean energy for M{\o}ller scattered electrons is
79\,MeV while the mean energy for elastic scattered electrons is 848\,MeV.
This offers the possibility of tuning the energy response of
the water Cherenkov luminosity detector to further suppress the unwanted
elastic scattering process off the proton relative to M{\o}ller scattering.
\begin{figure}[htbp]
  \begin{center}
    \epsfig{file=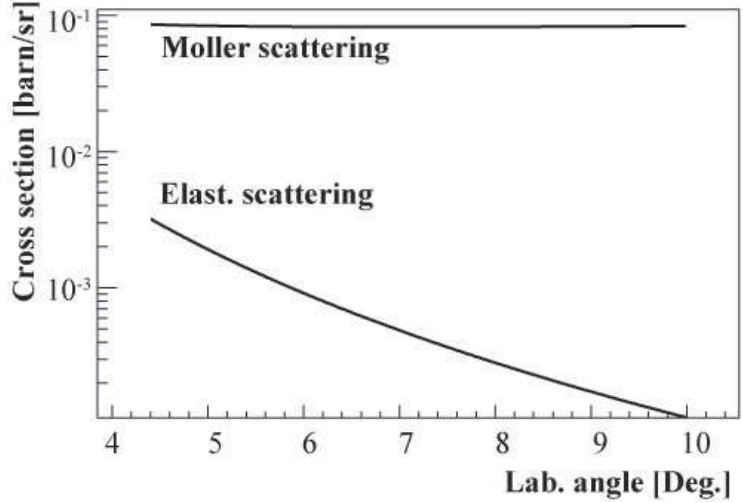,width=0.7\textwidth}
    \caption{Differential cross section of M{\o}ller scattering and elastic
    scattering as a function of the electron scattering angle in the
    laboratory $\theta_e$.
    The water Cherenkov luminosity detector spans an interval in $\theta_e$
    of $4.4^\circ < \theta_e < 10^\circ$.}
    \label{fig:wqmoellerunda0}
  \end{center}
\end{figure}
Elastic scattering on protons and M{\o}ller scattering dominate
at small forward angles between $4.4^\circ < \theta_e < 10^\circ$.
As shown in Fig.~\ref{fig:wqmoellerunda0}, the probability for M{\o}ller scattering of beam
electrons on electrons of the hydrogen target is a factor
of 100 larger as compared to elastic electron proton scattering.
The differential cross section for M{\o}ller scattering
is almost constant over a wide range of scattering angle while the elastic
cross section falls off as a function of the scattering angle.

An additional design requirement came from the fact that losses in counting rate
due to dead time in the detector or the experiment electronics
of the parity violation experiment have to be considered.
The losses due to double hits affect the measurement in two ways:
on one hand, pile-up reduces the measured counting rate, which can
be compensated by an extension of the measuring time. On the other hand,
pile-up losses due to background processes having a polarization dependent cross section can
lead to a systematic change of the measured asymmetry. The first and most important
step for reducing dead time effects was to find a detector material, that
has a fast response behavior and, therefore, reduces dead time effects to the per cent level.
The PbF$_2$ material employed in the A4 calorimeter is an
intrinsically fast Cherenkov radiator with no slow
scintillation components~\cite{pbf2:achenbach:01}.
The PbF$_2$-detector electronics measures the
energy within a 20\,ns integration gate, which is the largest dead time
in the system. It has a dead time-free digitization and a double hit (pile-up)
identification for intervals of more than 5\,ns~\cite{Koebis}.
Due to the non-linearity of the double hit rate losses, the fluctuations in the luminosity
have to be measured separately for both polarization states to be able to correct
for systematic changes in the experimental asymmetry. A detailed analysis
of the effect of pile-up losses from a polarization dependent background process
reveals a dependence on four quantities \cite{Hammel:dis:2004}:
\begin{enumerate}
\item The mean pile-up probability  $\lambda$. Even with a perfect
nonfluctuating beam where the polarization is equal for both helicity
states, a finite pile-up probability leads to a systematic change of the
measured asymmetry.

\item A difference of the pile-up probabilities
$\Delta \lambda = \lambda^+ - \lambda^-$ for the different helicity states of
the beam arises if the luminosity for the two helicity states is different:
$\langle \mathcal{L}^+ \rangle \neq \langle \mathcal{L}^- \rangle$.

\item Fluctuations in the pile-up losses caused by fluctuations in
the luminosity:
$\eta (\frac{\delta \mathcal{L}^{+^{2}}}{\langle \mathcal{L}^+ \rangle}+\frac{\delta
\mathcal{L}^{-^{2}}}{\langle \mathcal{L}^- \rangle})$, where
$\eta = 2 \tau \Sigma_0$, with $\tau$ the dead time (20~ns) and $\Sigma_0$
the cross section for elastic scattering. $\delta
\mathcal{L}^{\pm}$ is the RMS of the luminosity. There is a
systematic change of the measured asymmetry even
if the integrated luminosities are equal, $\langle \mathcal{L}^+ \rangle = \langle \mathcal{L}^- \rangle$,
and even if the beam fluctuations are symmetric in helicity, $(\delta \mathcal{L}^+)^2=(\delta \mathcal{L}^-)^2$.

\item The difference of the pile-up losses caused by asymmetric luminosity fluctuations
$\eta (\frac{\delta \mathcal{L}^{+^{2}}}{\langle
\mathcal{L}^+ \rangle}-\frac{\delta \mathcal{L}^{-^{2}}}{\langle
\mathcal{L}^- \rangle})$. It causes a systematic change of the
measured asymmetry if either the integrated luminosity or the RMS of the luminosity
is different for the different helicity states, $(\delta \mathcal{L}^+)^2 \neq (\delta \mathcal{L}^-)^2$.
\end{enumerate}

We obtain the following formula for the systematic change of the measured asymmetry depending on
the quantities defined above (please note that for the discussion of pile-up effects, we neglect
for a moment other apparative asymmetries like A$_I$ or A$_L$. If they are taken
into account, we yield A$_{mathrm{Exp}}$ from A$_{\mathrm{Meas}}$):
\begin{eqnarray}
     A_{\mathrm{Meas}} \! = \! \frac{
               \!-\! \Delta \lambda \!+\! (A_u \!+\! A_0) 2 \lambda P \!+\! 2 A_0 P
               \!-\! \eta \! \left( \! \frac{\delta \mathcal{L}^{+^{2}}}{\langle \mathcal{L}^+ \rangle}
               \!-\! \frac{\delta \mathcal{L}^{-^{2}}}{\langle \mathcal{L}^- \rangle} \! \right)
               \!+\! \eta (A_u \!+\! A_0) \! \left( \! \frac{\delta \mathcal{L}^{+^{2}}}{\langle \mathcal{L}^+ \rangle}
               \!+\! \frac{\delta \mathcal{L}^{-^{2}}}{\langle \mathcal{L}^- \rangle} \! \right) \! P
                    }
                    {
               2 \!-\! 2 \lambda \!-\! (A_u \!+\! A_0) \Delta \lambda P 
               \!-\! \eta \! \left( \! \frac{\delta \mathcal{L}^{+^{2}}}{\langle \mathcal{L}^+ \rangle}
               \!+\! \frac{\delta \mathcal{L}^{-^{2}}}{\langle \mathcal{L}^- \rangle} \! \right)
               \!-\! \eta (A_u \!+\! A_0) \! \left( \! \frac{\delta \mathcal{L}^{+^{2}}}{\langle \mathcal{L}^+ \rangle}
               \!-\! \frac{\delta \mathcal{L}^{-^{2}}}{\langle \mathcal{L}^- \rangle} \! \right) \! P
                    } \nonumber \\
 \label{eq:9quadpile}
\end{eqnarray}
Even for very symmetric beam conditions with equal integrated luminosities
$\langle \mathcal{L}^+ \rangle = \langle \mathcal{L}^- \rangle$, equal
luminosity fluctuations $ \delta \mathcal{L}^{+^{2}} = \delta \mathcal{L}^{-^{2}}$, and
equal polarization for both helicity states $ P^+ = P^-$, double hit losses from a polarization dependent
background process with asymmetry $A_u$ lead to a systematic change of the measured asymmetry $A_{\mathrm{Meas}}$
as compared to the physics asymmetry $A_0$.
In this case Eq.~\ref{eq:9quadpile} may be simplified to:
\begin{eqnarray}
  A_{\mathrm{Meas}} = \frac{A_0 P - (A_u + A_0) \lambda P}{1 - \lambda}
  \label{eq:exakt}
\end{eqnarray}
Due to rate losses by pile-up, as just discussed, an important design requirement
was the measurement of the variance (RMS) of the luminosity signal, in addition to
the integrated luminosity  $\langle L
\rangle$, by measuring the second moment of the luminosity, $\langle
L^2 \rangle$. From this, the variance can be extracted: $\langle
( \langle L \rangle - L )^2 \rangle = \delta L^2 = \langle L^2 \rangle
- \langle L \rangle^2$. The required accuracy of the luminosity measurement
for an experimental error of $\delta$A$_{\mathrm{Exp}}$/A$_{\mathrm{Exp}}$ = 1~\%
in a measurement gate of 20\,ms is:
\begin{eqnarray}
 \frac{\Delta L}{L} = 3 \times 10^{-5}, \quad
 \sqrt{\frac{\delta L^2}{\langle L \rangle ^2}} = 5 \times 10^{-3}.
\end{eqnarray}
This ensures that the experimental asymmetry can be corrected to
pile-up effects caused by luminosity fluctuations with an
accuracy of better than 1\%.
On account of the short dead time of the PbF$_2$ detectors and the fast
readout electronics, the losses due to double hits in the calorimeter are
1.7\% at 20\,$\mu$A beam current and a luminosity of 5~$\times$~10$^{37}$~cm$^{-2}$s$^{-1}$.

False asymmetries may also arise from noise signals correlated with
the power line frequency (50\,Hz) and higher harmonics as well as
beam fluctuations or systematic effects correlated with the helicity flip.
For this reason, the measurement of the luminosity is synchronized to the power line
frequency, i.e. the integration gate length is the inverse of the
current power line frequency ($\approx$20\,ms).
The length of the integration gate is synchronized by a phase locked loop
(PLL) to the line frequency.
For normalization purposes the gate length was measured for each helicity state.
Between two 20\,ms integration gates a 80\,$\mu$s time window for the change
of the electron beam helicity by changing the high voltage at the Pockels
cell~\cite{polsource:aulenbacher:97} is needed.
The 80\,$\mu$s time window guarantees that the Pockels cell
voltage has reached a stable condition. However, it introduces a phase shift of
the integration gate with respect to the line frequency. The phase shift
is (80/2080)2$\pi$ and after 25 gate pulses the phase difference is zero again.

We use two different patterns of helicity for four consecutive
integration gates: ($+ - - +$) and its complement ($- + + -$). Both
have the same number of $+$ and $-$ helicity states.
The patterns are chosen randomly by a bit shift register. Since a helicity flip corresponds
to a fast high voltage change at the Pockels cell which might induce
electromagnetic pick-up signals the patterns chosen guarantee that the helicity
flip is equiprobable and that a correlation of
the asymmetry with the polarization sequence is avoided.

\section{Design and construction of the luminosity monitor}
The design of the luminosity monitors is based on a water Cherenkov detector system
read out by photomultiplier tubes. Eight
modules are used which cover an azimuthal electron scattering angle $\phi_e$
of 2$\pi$, while the polar electron scattering angle range is $4.4^\circ < \theta_e < 10^{\circ}$.
The high event rate of 5.5~GHz per module requires an integrating
measurement of the anode current of the photomultiplier tubes.

\begin{figure}[htbp]
  \begin{center}
    \epsfig{file=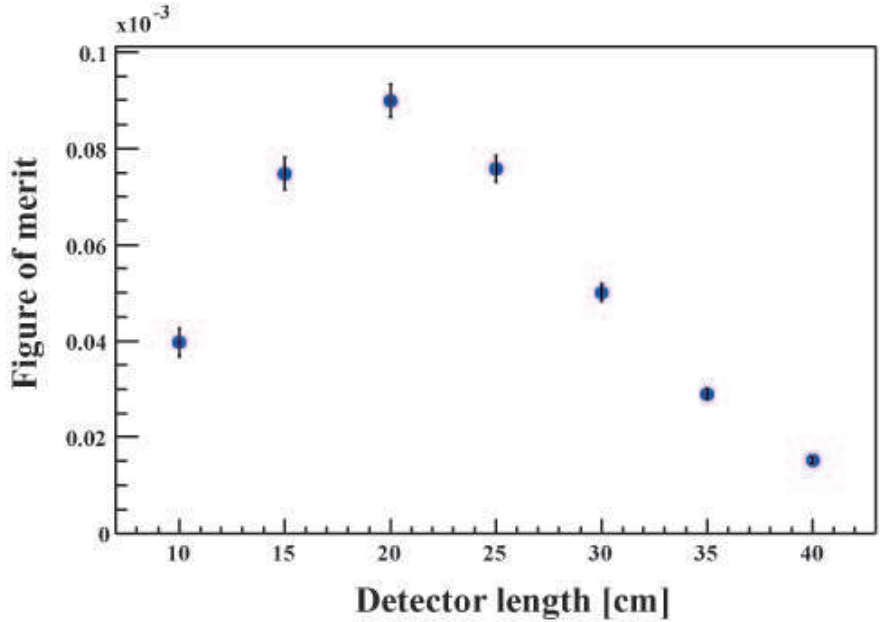, width= 0.9 \textwidth}
    \caption{Figure of merit (as defined in Eq.~\ref{eq:fomlumi}) as a function of the
    length of the water volume determined by GEANT simulations (see text).}
    \label{fig:fom}
  \end{center}
\end{figure}

\begin{figure}[htbp]
  \begin{center}
    \begin{tabular}{l}
      \epsfxsize=3.5in\epsfbox{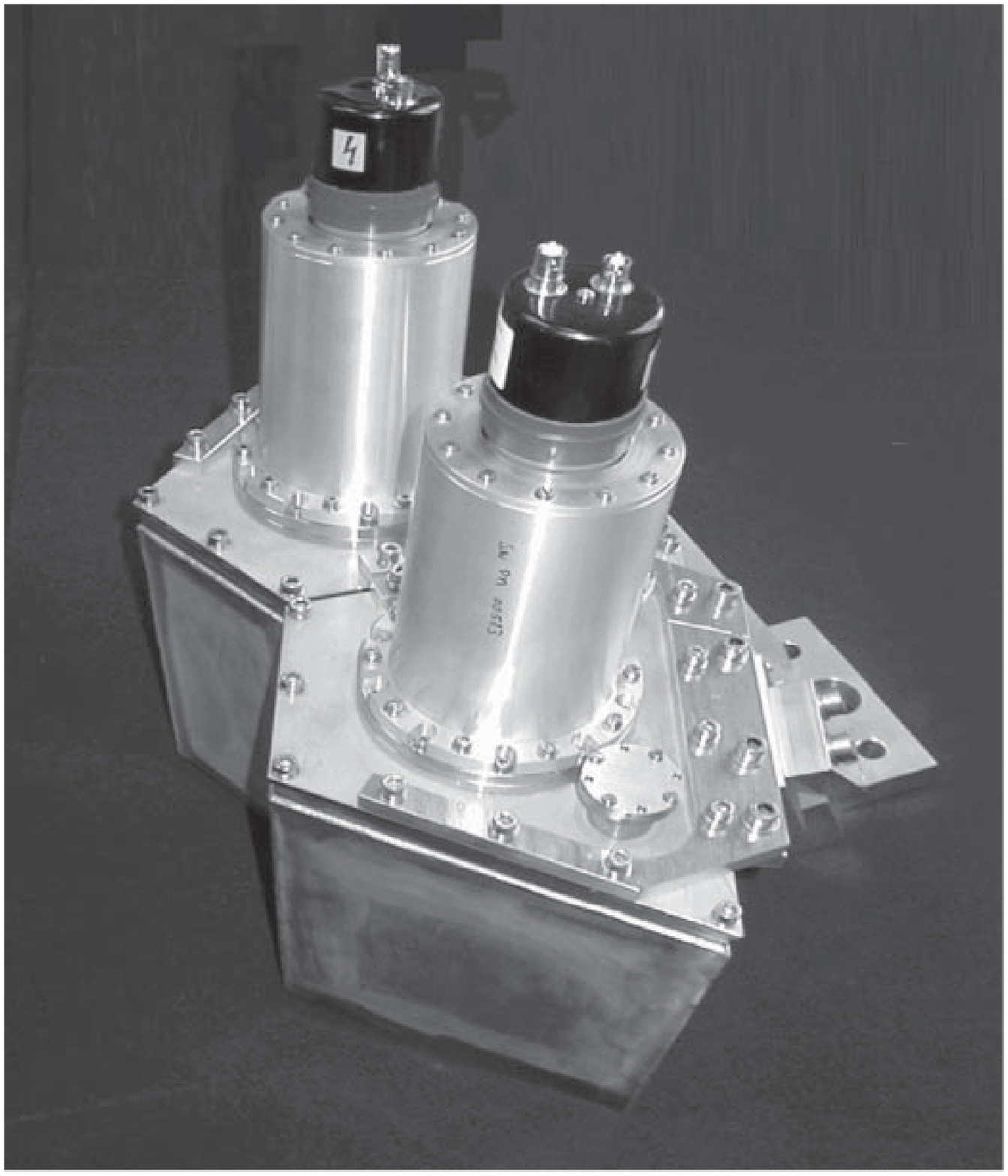} \\[0.5cm]
      \epsfxsize=4.5in\epsfbox{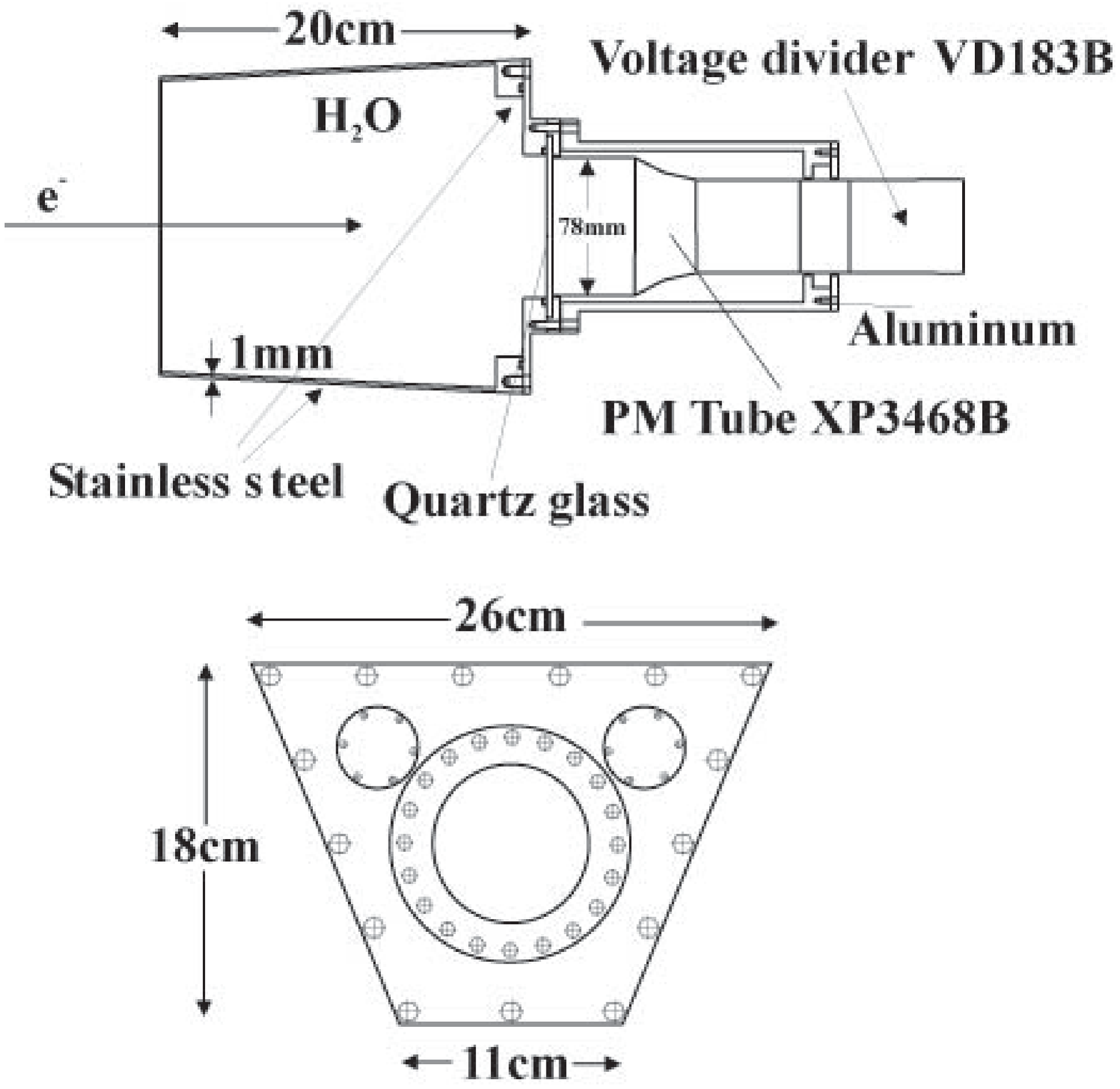}
    \end{tabular}
    \caption{Top: Photograph of two out of eight neighboring luminosity monitors
      with the photomultiplier tubes attached. Bottom: Detailed CAD drawing of a
      luminosity monitor in side view and from the rear face.}
    \label{fig:lumopics}
  \end{center}
\end{figure}

With the program package GEANT 3.21 of the CERN software library the
response behavior of the luminosity was simulated and
optimized. As already discussed, the energy of M{\o}ller electrons is
an order of magnitude lower than that of elastic scattered electrons.
The simulations were done for a stainless steel tank
filled with demineralized water and a photomultiplier tube attached to
a quartz-window of the steel tank.
The optical transmission of water was taken into account as well as the
quantum efficiency of the photocathode. In the simulations we selected
rhodium mirrors on the inner faces of the luminosity monitor.
Despite its high price, rhodium was chosen due to its reflectivity
and good resistance to ionized water.
The energy loss of electrons in the water volume,
the production of Cherenkov light, and its collection
on the photocathode of the photomultiplier tube was
simulated. Electrons with appropriate energies were injected
directly into a module of the luminosity monitor.
Three parameters turned out to be the most important for
the response behavior of the detector:

\begin{itemize}

\item[1.] The first variable is the mean number of
Cherenkov photo-electrons from M{\o}ller electrons $N_{\mathrm{M{\o}ller}}^{Photons}$
which determines the statistical accuracy of the luminosity measurement.
It is given by the development of the electromagnetic shower
induced by a M{\o}ller electron as it evolves in the water volume
and also by the production of Cherenkov photons
from charged particle tracks.
\item[2.] The fluctuation of $N_{\mathrm{M{\o}ller}}^{Photons}$
has two contributions: the usual $\sqrt{N}$ behavior and an
additional fluctuation coming from the electromagnetic shower fluctuations.
In order to minimize the shower fluctuations we optimized the ratio
$E^\mathrm{M{\o}ller}/\sigma_E^{\mathrm{M{\o}ller}}$ of the
mean deposited energy $E^\mathrm{M{\o}ller}$ of M{\o}ller electrons in the detector volume
and the RMS of the energy deposit of M{\o}ller electrons
$\sigma_E^{\mathrm{M{\o}ller}}$.
\item[3.] In addition, we wanted to suppress the response of the detector
to elastic scattered electrons by maximizing the ratio of the
mean deposited energy of the M{\o}ller electrons
over that of elastic scattered electrons $E^{\mathrm{M{\o}ller}}/E^{\mathrm{elastic}}$.
\end{itemize}
For determining the design parameters of luminosity monitors
we have simulated different lengths of the detector modules in
search of the maximum of the product of the three factors discussed above:
\begin{eqnarray}
   \mathrm{FOM} &=&  N_{\mathrm{M{\o}ller}}^{Photons} \frac{E^\mathrm{M{\o}ller}}{\sigma_E^{\mathrm{M{\o}ller}}} \frac{E^{\mathrm{M{\o}ller}}}{E^{\mathrm{elastic}}}\\
  \label{eq:fomlumi}
\end{eqnarray}
The figure of merit as extracted from the full GEANT simulation of the
water Cherenkov luminosity detector is shown in Fig.~\ref{fig:fom}.
The optimum detector length can be read off the figure to be 20\,cm.
Due to the ring geometry the monitor modules have the shape of a
frustum of pyramid with a trapezoidal base. The depth of a module in
beam direction is 20\,cm. The readout of Cherenkov light is done
with quartz window photomultiplier tubes which are separated by quartz windows
(HOQ310, Heraeus Quartzglas, Hanau, Germany) from the water volume and
are placed in a light-tight aluminum housing (Fig.~\ref{fig:lumopics} top). The luminosity
monitors are attached to the downstream end of the A4 scattering
chamber. The flange of the scattering chamber was milled down to 5~mm thickness
in the angular range from $4^\circ < \theta_e < 10^\circ$ in order to minimize the energy loss
of scattered electrons before entering the luminosity monitors.

Due to the high radiation dose during the experiment of $>1$\,Mrad per 1000~h
it
is necessary to use radiation resistant materials. The photomultiplier
tubes are equipped with quartz windows. In order to avoid the
production of radicals in the water volume distilled water is
used. The detector modules had to be built with as little
material as possible in order to minimize backscattering from the
luminosity monitors into the PbF$_2$ calorimeter.
The material of the support structure of the monitors had been minimized,
too. Each of the eight modules
consists of a water tank with 1\,mm thick steel walls. Two of the modules
are shown in Fig.~\ref{fig:lumopics}. The
detector housing is manufactured of high-grade stainless steel in
order to avoid corrosion by the water. The photomultiplier tube,
attached to the rear face of the water tank, is housed in a light tight
cylinder made of aluminum. In the experiment, we did not use
rhodium mirrors at the inner faces.
The voltage divider of the photomultiplier tube is outside
of this cylinder to avoid overheating.
The two flanges on the rear face of the water tank would allow the circulation of the water,
if necessary. During our experiment we found
that this was not necessary. Due to the activation during the experiment all parts of the
detector are fixed by screws so that in case of a defect a module can be
dismounted within a very short time.
Figure~\ref{fig:lumopics} (bottom) shows a drawing
with details of the detector. Table~\ref{tab:LUMOdata}
summarizes the specifications of the luminosity monitors.
\begin{table}[htpb]
    \caption{Specifications of the water Cherenkov luminosity detectors:   \label{tab:LUMOdata}}
    \begin{center}
    \begin{tabular}{| l | r |}  \hline
    \multicolumn{2}{| c |}{water Cherenkov detector optimized for M{\o}ller (FOM)}\\ \hline
    8 modules                        & $4.4^\circ < \theta_e< 10^\circ$ \\
                                     & $0^\circ   < \phi_e  < 360^\circ$\\
                                     & $\Delta \phi_e = 45^\circ$       \\ \hline
    solid angle per module           & 9.97 msr                        \\ \hline
    rates per module:                &  0.3~GHz (elastic)              \\
                                     & 43.8~GHz (M{\o}ller)            \\
                                     & 44.1\,GHz total event rate      \\ \hline
    readout                          & integrating anode current        \\ \hline
    accuracy per module:             & $\Delta \mathcal{L}/\mathcal{L} = 3.4 \cdot 10^{-5}$ (in 20\,ms)  \\ \hline
    radiation dose                   & $\approx$ 1.4\,Mrad per 1000~h     \\ \hline
    \end{tabular}
    \end{center}
\end{table}

We use Philips XP3468B photomultiplier tubes with 8 dynodes
where we have bypassed the last three dynodes. This
photomultiplier tube has a diameter of 76\,mm and a length of
164\,mm. The entrance window has a diameter of 76\,mm, the sensitive
surface of the photocathode has a diameter of 68\,mm. This tube size
was selected for a good coverage of the rear face of the luminosity
monitors. A transparent coupling in the whole spectral region must be
achieved at the connection between the detector window and the photomultiplier tube,
for which the radiation resistant silicone rubber Elastosil RT 601 is
used. The rate per module amounts to $44.1 \cdot 10^9$\,Hz. One
M{\o}ller scattered  electron produces approximately 170 photons. Per 20~ms
integration gate, $7.5 \cdot 10^{12}$ photons are produced. With a quantum
efficiency of the photocathode of about 20\% this corresponds to $1.5
\cdot 10^{12}$ photoelectrons. The maximum input
current of the electronics amounts to 100~$\mu$A equivalent to $6.24
\cdot 10^{14}$ electrons per second. Accordingly, the amplification factor of
the photomultiplier tube has to be limited to $5 \cdot 10^3$
which was realized by using only 5 dynodes.

\section{Electronics and Readout}
The A4 experiment luminosity electronics was developed and built in-house.
For the eight signals from the luminosity monitors,
the raw signal and the squared signal have to be integrated
and digitized. There are more signals from beam parameter monitors like horizontal
and vertical position monitors at different places in the accelerator,
electron beam current monitor, and electron energy monitor in the accelerator etc.
These signals are integrated over 20~ms, digitized and histogramming by modules
of this part of the electronics.
The A4 PV-experiment is controlled by a gate generator which delivers
a gate signal to the calorimeter electronics and monitoring electronics.

\begin{figure}[htbp]
  \begin{center}
    \epsfig{file=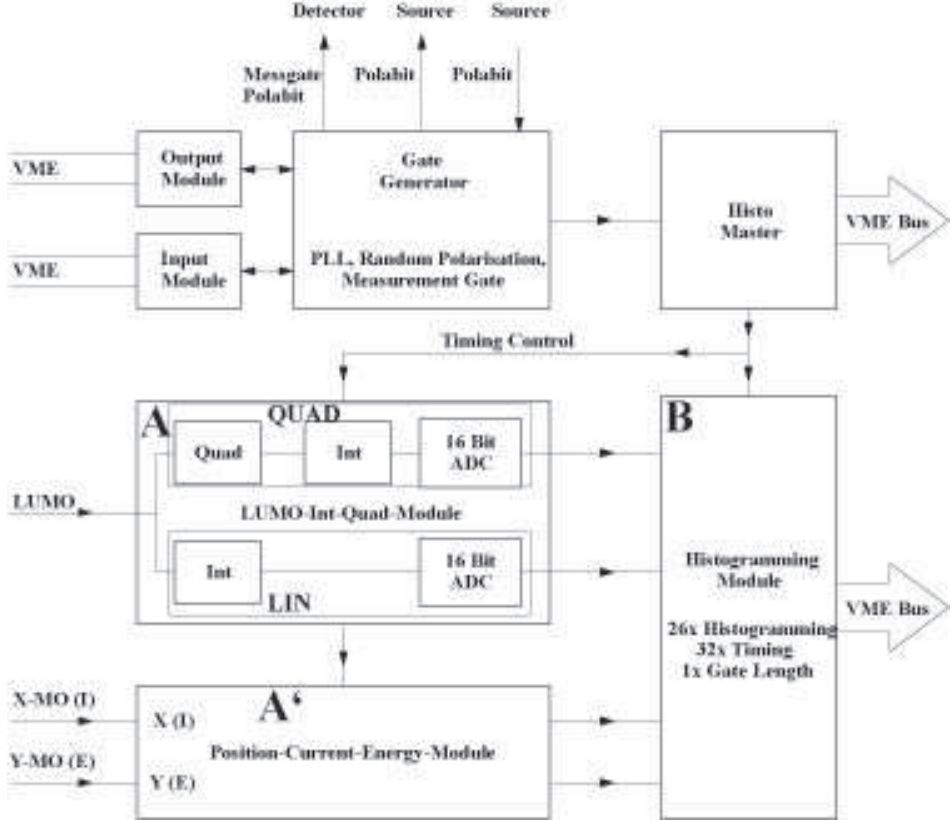, width= 0.9 \textwidth}
    \caption{Scheme of the A4 monitoring electronics.}
    \label{fig:electronics}
  \end{center}
\end{figure}

An electronics unit for one water Cherenkov luminosity monitor
consists of two modules (module A and module B in Fig.~\ref{fig:electronics}).
Module A contains two internal channels (LIN and QUAD in Fig.~\ref{fig:electronics}).
Channel LIN  contains an integrator circuit for the linear luminosity signal and
a 16 bit analog-to-digital (ADC). Channel QUAD has, in addition, a squaring device
in front of the integration circuit. The squaring device is an analogue multiplication
device. The accuracy of the linearity of the electronic multipliers amounts
to $\approx 5\cdot10^{-3}$. The integral and
differential linearity of the 16\,bit ADC is less than
1\,LSB (least significant bit). The module A with the integrator (LIN)
and squaring device (QUAD) permits
a measurement of the first and second moment of the luminosity.
The module B contains the histogramming modules.
Data are transferred by an optically insulated serial interface
to the histogramming units. The
histogramming module has 4\,MB RAM per channel (16\,bit data depth,
21\,bit addresses) which can store separated histograms for each polarization state.
The integration gate of
20\,ms can be divided, if necessary, into 16 time slices, so that the
structure of the luminosity within a 20\,ms gate can be examined. One
time slice has a length of 1.25\,ms.

The measurement of small (of order $10^{-6}$) asymmetries requires special care
in isolating the integrator, squaring device, and ADC - all housed in module A
of Fig.~\ref{fig:electronics} - from outside
noise or electromagnetic pickup.
The module A and B are galvanically separated so that the digital signals from the
histogramming module B can not induce noise in module A.
The polarization information (polarization bit, $P$-bit) is electronically encoded
in different voltage and current levels which could cause small changes of the current and voltage
distribution of the electronic circuit. This results in different current distributions
on the ground plane of the electronics circuit and different potential differences
for the two helicity states.
A careful design of the ground plane in the electronics circuit of module A
is necessary in order to avoid small shifts of the zero Volt ground level at the
input of the integrator, squaring device, or ADC which would result in
false asymmetries of the measured signal charges. Such cross talk of the P-bit
to the luminosity measurement is carefully avoided.

The histogramming module B contains a programmable logic unit which allows
to operate the module in different ways:
\begin{itemize}
\item[-] If operated as a pure histogramming module, the output of the ADC
        is - depending on the value of the $P$-bit - interpreted as an address
        and the memory content of that address cell
        is increased by one. After the run measurement time one can readout
        two histograms (one for each polarization state) directly via VME-bus
        from the memory of the histogramming module. Correlation between different
        luminosity monitor modules is lost that way but has the advantage that
        no additional data treatment is necessary to create histograms out of the data.
\item[-] For diagnostics reasons one can operate the module in a \lq\lq timing\rq\rq\
        mode, where each 20~ms integration the ADC-output is stored directly in memory.
        This mode stores all signals as a function of time up to 30 minutes and gives the opportunity
        to correlate all signals with each other. The polarization bit is stored with
        each ADC value.
\end{itemize}
The timing control of the integrators, ADCs
and the histogramming units is synchronized by one additional module
(\lq\lq Histo Master\rq\rq\ in Fig.~\ref{fig:electronics}) giving control signals
to all the modules.
The noise of the electronics and any false asymmetry induced by the electronics
was measured to be well below the desired accuracy of $3 \cdot 10^{-5}$
and can, therefore, be safely neglected.

Additional beam monitor signals from the accelerator are integrated and digitized by a module similar
to module A in Fig.~\ref{fig:electronics}. This additional module
(Module A' in Fig.~\ref{fig:electronics}) has two individual input channels with a
linear integrator and ADC each. In total, one has 8 Luminosity monitor signals
(LIN and QUAD in 8 modules from type A) and 10 more signals from the
accelerator (readout by 5 modules from type A').
The data from the modules of type A and A' are stored
twice in two modules from type B in order to create redundancy and detect changes
of individual hardware memory bits by ionizing radiation.

A further VME-bus module measures
the length of the integration gate for each polarization state by
a 35\,bit counter and histograms it. An external 20\,MHz quartz (generator)
serves as time base. The measured asymmetry in the integration gate length,
which serves as the integration and counting window for the whole parity experiment,
is less than 10$^-9$.
Thus, a high absolute accuracy for the measurement of small
(order 10$^{-6}$) asymmetries in the luminosity and other
accelerator beam parameter signals was achieved.

\section{Performance and Measurements}
\begin{figure}[htbp]
  \begin{center}
    \epsfig{file=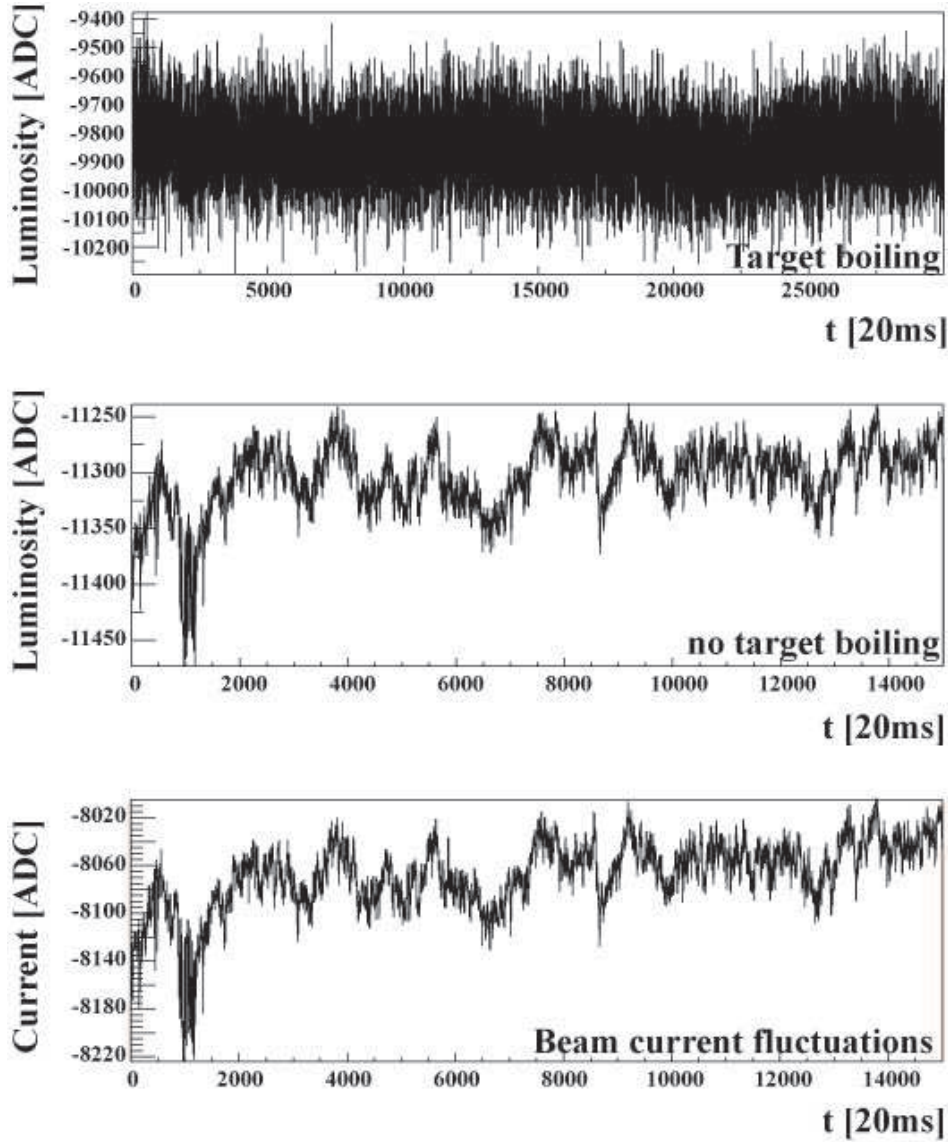, width= 0.9 \textwidth}
    \caption{Example of target density and beam current fluctuations.}
    \label{fig:lumidensifluc}
  \end{center}
\end{figure}

The performance and accuracy of the luminosity monitors was
investigated in several experiments. With the luminosity monitors, the
target density fluctuations of the liquid hydrogen target were studied.\\
From a simultaneous measurement of the beam intensity and the luminosity,
target density fluctuations can be detected and analyzed.
The result of such a study is shown in Fig.\ref{fig:lumidensifluc},
for different conditions of the electron beam size and position.
In Fig.~\ref{fig:lumidensifluc} the luminosity monitor (LUMO) signal and the
signal of a beam current monitor (PIMO) are shown as a function of time.
The PIMO is a microwave cavity with high Q-value which can be
adjusted to measure either the phase (P) or the current (I) of the electron beam.
For Fig.~\ref{fig:lumidensifluc} the PIMO has been adjusted to measure
the electron beam current.
The top plot shows
a signal of a single module as a function of time sampled every 20~ms, where the target
has been operated in such a way so that target density fluctuations from boiling
dominate. Boiling causes a fluctuation in the luminosity signals of $\pm$2.5\%.
For the middle figure, the diameter of the electron beam and the cooling performance
of the hydrogen target were optimized and the beam had been 1~mm off axis.
The fluctuations of the measured luminosity was reduced to $\pm$0.5\%
which can be identified as beam current fluctuations comparing them
to the independent beam current signal in the bottom plot.

\begin{figure}[htbp]
  \begin{center}
    \epsfig{file=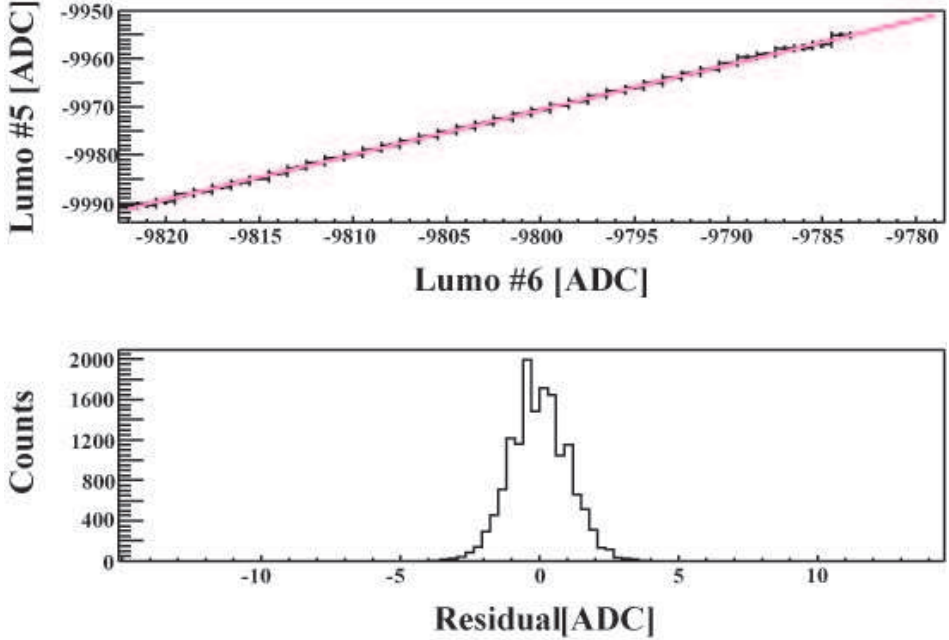, width= 0.9 \textwidth}
    \caption{Determination of the accuracy of the luminosity
    monitors.}
    \label{fig:lumoaccuracy}
  \end{center}
\end{figure}

The correlation of the signals of two luminosity monitors allows us
to separate intrinsic detector fluctuations like uncorrelated noise
from external fluctuations in the luminosity signal caused by
beam or target density fluctuations. The latter are correlated
for two neighboring modules. For the determination of the uncorrelated noise,
which limits the sensitivity, two
neighboring luminosity monitors were combined. For two adjacent neighbors helicity
correlated differences in position and angle are small.
In Fig.~\ref{fig:lumoaccuracy}, the averaged signal of luminosity monitor \#5 is plotted versus the
averaged signal of luminosity monitor \#6.
Both signals are strongly linearly correlated. We conclude that
the observed signal fluctuations are caused by target density and beam
fluctuations. For the graph in Fig.~\ref{fig:lumoaccuracy} (bottom)
the residual of the non-averaged luminosity data of the same monitor and the straight line fit
was calculated and sorted into a histogram. The RMS width of this distribution divided by $\sqrt{2}$
gives the experimental accuracy of a single luminosity measurement in 20\,ms, $\delta L_{20ms}$.
The RMS width divided by $\sqrt{N}\sqrt{2}$ gives the accuracy for our
data taking running time of 5 minutes,
$\delta L_{5min}$. For the data in Fig.~\ref{fig:lumoaccuracy},
the mean luminosity signal corresponds to $(9862.4 \pm 0.006)$\,ADC
channels. The relative accuracy results  $(\delta
L/L)_{5min} = 6.1\cdot10^{-7}$ in 5 minutes and $(\delta L/L)_{20ms} =
7.5\cdot10^{-5}$ in 20\,ms. In comparison with the simulated accuracy
of $(\Delta L/L)=3 \times 10^{-5}$ it is worsened by a factor 2-3. This
can be explained by the fact that the luminosity monitors are used
without rhodium mirrors with correspondingly less light reaching the photocathode
of the photomultiplier tube. Tests have shown that mirrors improve the
accuracy by a factor 3-5. Thus, we can conclude that the simulations
and the experimental results from the real luminosity monitors
are in good agreement.

\begin{figure}[htbp]
  \begin{center}
    \epsfig{file=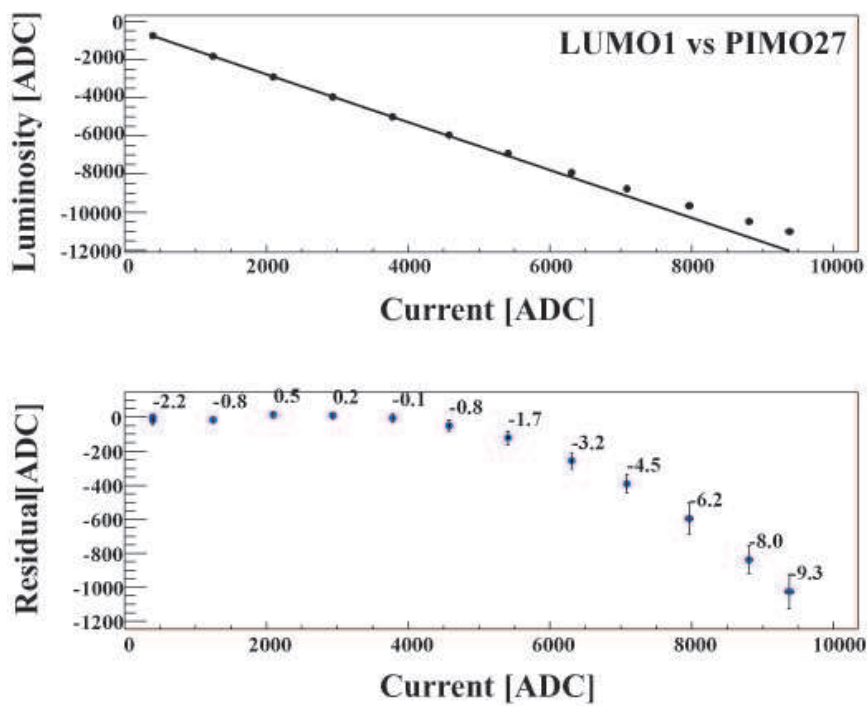, width= 0.9 \textwidth}
    \caption{Top: Signal of luminosity monitor as a function of the
    beam current monitor signal. Bottom: The points give the absolute
    difference of the measured data from the
    the straight line fit. The numbers give the relative deviation in percent.}
    \label{fig:linearityLUMO}
  \end{center}
\end{figure}

A correction was applied for the non-linearity of the luminosity
monitor photomultiplier tubes. This was measured and verified
separately by varying the beam current from 0 to 23\,$\mu$A.
A false asymmetry results from non-linearities in the response
behavior of the luminosity system. The nonlinearity has to be measured
and the data have to be corrected, accordingly. Without this correction
the false asymmetry would be about $5\times10^{-9}$ per percent
nonlinearity (see Fig.~\ref{fig:linearityLUMO}) at a
beam current asymmetry of 10${-6}$.
A non-linearity of 1~\% in the response behavior of the luminosity system
causes a false asymmetry of $5\cdot10^{-9}$.
Fig.~\ref{fig:linearityLUMO} shows the signal of a luminosity
monitor as a function of a beam current variation from 0 to 23\,$\mu A$
as measured with the linear beam current monitor PIMO.
Nonlinearities at high beam currents are readily seen in the upper plot.
The lower part shows the local steepness by deviation from
zero. The steepness varies as much as 8~\% from 0~$\mu$A to 20~$\mu$A.
The non-linear deviation from the straight line fit is 8.0\% at the
operating point of the A4 experiment of 20\,$\mu$A (here the PIMO
signal is $\approx$~8700 ADC channels). If not corrected this would
cause a false asymmetry of $-4 \cdot 10^{-8}$ for a beam current asymmetry of $1
\cdot 10^{-6}$. Actual beam current asymmetries vary between $1 \cdot
10^{-8}$ and $1 \cdot 10^{-5}$.

\begin{figure}[htbp]
  \begin{center}
    \epsfig{file=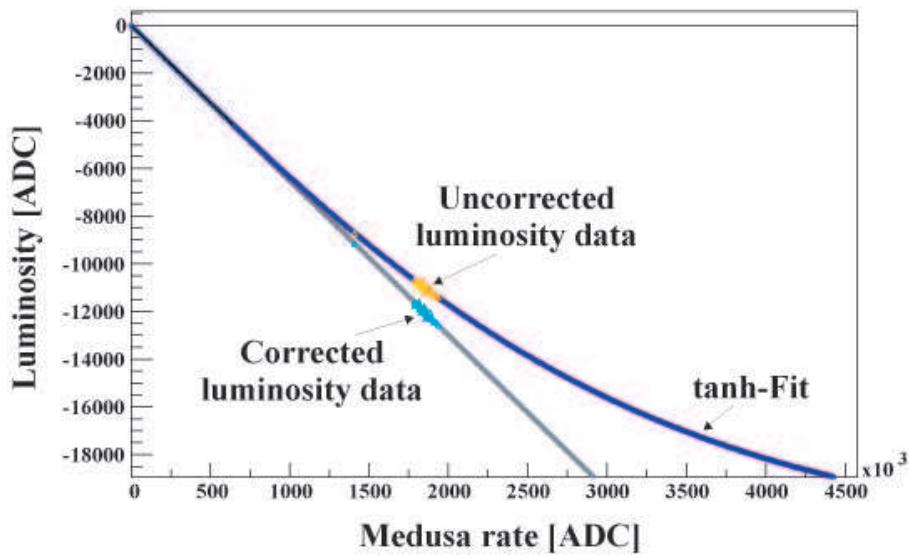, width= 0.9 \textwidth}
    \caption{Correction of LUMO-Data via hyperbolic tangent.}
    \label{fig:korrtanh}
  \end{center}
\end{figure}

We corrected the non-linearities in the luminosity signals with the
measured counting rates of the PbF$_2$ calorimeter since both
detector systems see the same luminosity. This detector
system was shown to be linear after correction for pile-up losses
even for high beam currents. The
non-linearity of the luminosity signals has been corrected without correlation
to helicity, i.e. the counting rates of both helicities of the luminosity
monitors and the counting rates of the PbF$_2$ calorimeter (in
Fig.~\ref{fig:korrtanh} called MEDUSA) were added.
Due to the small current asymmetry
within the range of $10^{-6}$ to $10^{-5}$ the correction factor of
the non-linearity of the luminosity monitors is not helicity dependent.
The interpolation of the non-linearity of the luminosity monitors
can be very well approximated by a $\tanh$-function.
Fig.~\ref{fig:korrtanh} shows the implementation of the
non-linearity correction. The figure shows, on the one hand, the luminosity
raw data on the tanh-curve and, on the other hand, the corrected
luminosity data on the straight line fit which describes the linearized and
corrected data.

\begin{figure}[htbp]
  \begin{center}
    \epsfig{file=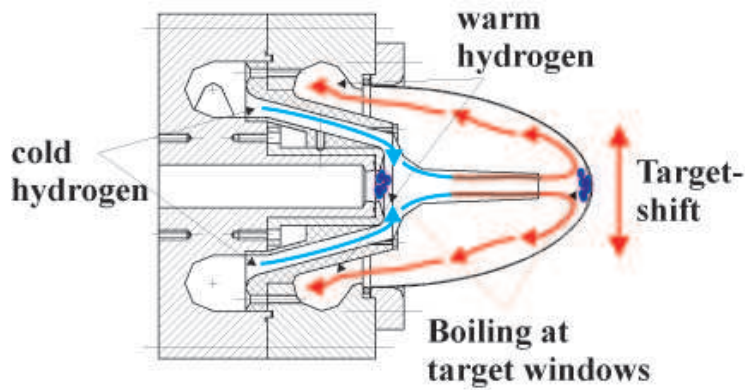, width= 0.7 \textwidth}
    \caption{Scheme of the target cell.}
    \label{fig:targetcell}
  \end{center}
\end{figure}

\begin{figure}[htbp]
  \begin{center}
    \epsfig{file=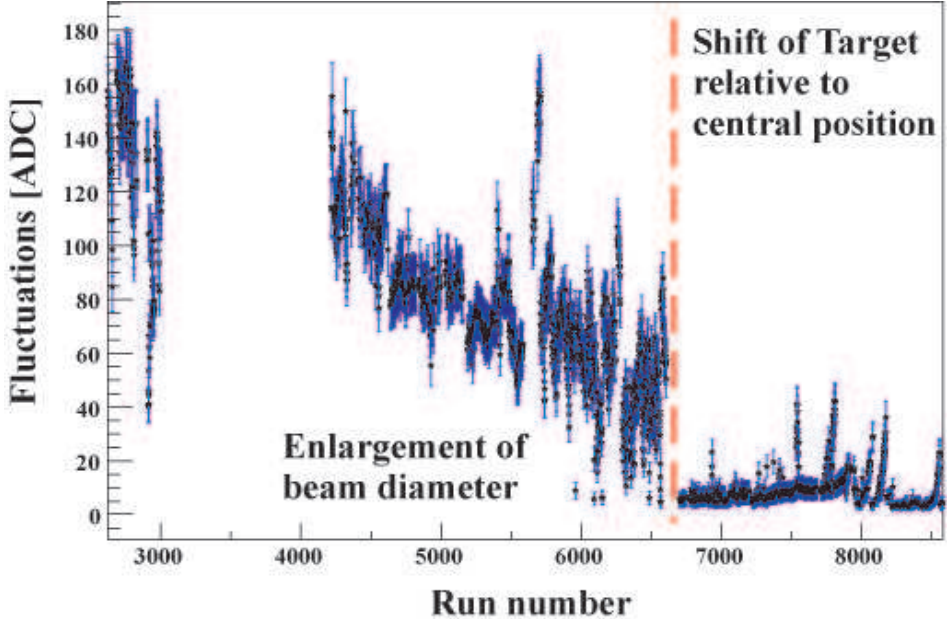, width= 0.9 \textwidth}
    \caption{Investigations of target density fluctuations.}
    \label{fig:targetboil}
  \end{center}
\end{figure}

Another important point is the investigation of target density
fluctuations caused by heating due to the energy deposition of
the electron beam in the hydrogen. The scheme of the target cell is shown in
Fig.~\ref{fig:targetcell}. The 10~cm liquid hydrogen target
provides a luminosity of $L = 5 \cdot 10^{37}$\,cm$^{-2}$s$^{-1}$ at
20\,$\mu$A electron beam current. This corresponds to about 100~W
of heat absorbed in the liquid hydrogen and in the two windows cooled by the hydrogen.
The target has a  closed loop circulating system cooled by a helium
refrigerator. Thus, the temperature of the hydrogen entering the target cell
is just above the hydrogen freezing point, much below the boiling temperature
(deeply subcooled hydrogen).
The design of the liquid hydrogen loop and the target cell were optimized
to obtain a high degree of turbulence with a Reynolds number of $R > 2 \times 10^5$
in the target cell in order to increase the effective heat transfer.
The heat exchange is intensified by transverse turbulent mixing
which causes a faster mass exchange across the hydrogen stream.
This approach removes the heat deposit by the electron beam which is
concentrated in a small range around the beam axis and which can cause
fast luminosity fluctuations from hydrogen
density variations or boiling of the liquid at the windows.
This new technique allowed us, for the first time, to avoid a fast modulation of the
beam position (rastering) of the intense CW 20\,$\mu$A electron beam. It
permitted us to stabilize the beam position on the target cell with less than
10$^{-3}$ relative target density fluctuations arising from boiling.
The cooling system supports a high flow rate of liquid
hydrogen and up to 250~W heat load of beam energy deposition in the
target.
We investigated the effect of beam diameter variations on target
boiling and luminosity fluctuations. In this context,
\lq\lq fluctuation\rq\rq\ refers to the point to point
difference of the measured luminosity signals for two neighboring
20\,ms integration gates. The fluctuations of the luminosity
signal were measured as a function of the electron beam cross section from
5$\times$~10$^5$~$\mu$m$^2$ up to 3$\times$~10$^6$~$\mu$m$^2$ and the impact
position of the electron beam at the target. The results of the measurements
are presented in Fig.~\ref{fig:targetboil}.
One recognizes that an enlargement of the beam diameter
(runs 3000-6500) causes a decrease of the fluctuations by a factor
4-5. The measured fluctuations correlate with increasing beam diameter as the
heat deposition in the target is distributed on a larger area and the
density of deposited energy is smaller.
Additional boiling at the entrance and exit windows of aluminum
has a large influence on the observed luminosity fluctuations.
The calculated specific heat flow from the windows into the liquid hydrogen
exceeds the critical point for boiling on the surface, even for an
enlarged beam spot.
Therefore, boiling at the windows does take place. A detailed analysis that
includes the aluminum heat conductivity shows that
this process does not depend very much neither on the beam size nor on the
wall thickness.
One way to reduce a bubble content in the beam
region is the increase of the mass flow tangentially to the window surface.
That is achieved in the present axially symmetric construction of the
target cell by shifting the target axis relative to the axis of the beam.
For a systematic investigation of the boiling at the target windows,
the target position was varied
in the range of 5 mm in vertical direction (from 2 mm below to 3 mm
above the beam position) and the fluctuations were
measured with the luminosity monitors.
It was found, that shifting the target position by 1 mm out of the
center reduces the fluctuations by a factor 4 (The result can be seen in
Fig.~\ref{fig:targetboil}, runs 6700-8500). If the electron beam hits
accurately the center of the target nose the heat cannot be removed
as efficiently as in a position 1 mm above or below the center of symmetry.
We found a
reduction in the width of the asymmetry distribution around an order
of magnitude after the reduction of the target boiling.
Eliminating the target boiling controlled by the luminosity monitor
plays an essential role in the experiment. The measured
asymmetry caused essentially by beam current asymmetries
reduces from $(-4.61 \pm 1.02)$~ppm to $(-0.89 \pm
0.31)$~ppm. The error of the measured asymmetry reduces from of
1.02~ppm to 0.31~ppm.

\section{Summary}
A luminosity monitor system was developed to correct the A4
asymmetry data to target density fluctuations.
The system determines the luminosity and the RMS of the luminosity
in a 20~ms integration gate. The system was studied
and optimized in simulations. The radiation-hard luminosity monitor
measures in pulse integrating mode at very small forward scattering
angles and achieves a relative accuracy of $\delta L/L =
7.5\cdot10^{-5} \pm 4\cdot10^{-7}$ in 20\,ms. The use of such a monitor
system not only allows us to learn about the state of the hydrogen target
but also enables us to correct the measured rates to the actual
(square) luminosity. This enhances the accuracy of the parity violation
measurement by a factor of 3.

\section{Acknowledgements}
This work was supported by the Deutsche Forschungsgemeinschaft
in the framework of the SFB 201 and the SPP 1034.
The development of this electronics took place in co-operation with
the electronics workshop of the Institut f{\"u}r Kernphysik Mainz
in collaboration with R.~B{\"o}hm, G.~Hacker, J.~Reinemann and H.~Streit.

\bibliographystyle{elsart-num}
\bibliography{nim_lumo_hammel}% Produces the bibliography via BibTeX.

\end{document}